\documentclass[preprint2]{aastex}

\slugcomment{}

\shorttitle{A search for radioactive $^{26}$Al in V4332 Sgr}
\shortauthors{AAAA et al.}

\begin{document}


\title{A search for radioactive $^{26}$Al in the nova-like variable V4332 
Sagittarii}


\author{Dipankar P.K. Banerjee}
\affil{Physical Research Laboratory, Navrangpura,  Ahmedabad 
Gujarat 380009, India}
\email{orion@prl.ernet.in}
\author{Nagarhalli M. Ashok}
\affil{Physical Research Laboratory, Navrangpura,  Ahmedabad 
Gujarat 380009, India}
\email{ashok@prl.ernet.in}
\author{Olli Launila}
\affil{SCFAB-KTH, Royal Institute of Technology, Atomic and Molecular Physics
 Department, Roslagstullsbacken 21, SE-106 91, Stockholm, Sweden}
\email{olli@physics.kth.se}
\author{Christopher J. Davis}
\affil{Joint Astronomy Center, 660 N. Aohoku Place, Hilo, Hawaii-96720, USA}
\email{c.davis@jach.hawaii.edu}
\and
\author{Watson P. Varricatt}
\affil{Joint Astronomy Center, 660 N. Aohoku Place, Hilo, Hawaii-96720, USA}
\email{w.varricatt@jach.hawaii.edu}


\begin{abstract}
We have searched for the  important radioactive isotope $^{26}$Al in the 
nova-like source V4332 Sgr. Recent results from gamma ray astronomy show 
that there is  pervasive emission of the 1.809 MeV gamma ray photon, 
arising from the decay of $^{26}$Al to $^{26}$Mg, from all over the 
galactic plane. Though the sites from where this emission originates are 
not clearly established,  novae are believed to be an important contributing
source. In this context, V4332 Sgr presented a rare opportunity
to observationally investigate  whether novae or novae-like sources 
synthesize $^{26}$Al and to what extent. Strong AlO bands in the near-IR  
have been reported in this object recently. As molecular bands of different
isotopic compositions are readily resolved spectroscopically (e.g. $^{12}$CO 
and $^{13}$CO), it was thought that the  components of AlO associated with
 $^{26}$Al  and stable $^{27}$Al could be detected as separate bands. 
Our spectra indicate that there is no strong presence of
$^{26}$Al in V4332 Sgr. A  reliable upper limit of 0.10 for the 
$^{26}$Al/$^{27}$Al ratio is determined  which constitutes
the first observational constraint for this ratio in a potential $^{26}$Al 
producing source. While V4332 Sgr is not
a typical nova, its outburst amplitude and light-curve behaviour bear close
similarity to that of novae. Hence, although the results from V4332 Sgr   
cannot be  directly extended to novae in general, the limit on the 
observed $^{26}$Al/$^{27}$Al ratio could  be a  useful input in constraining 
rather uncertain nucleosynthesis models  for the production 
of $^{26}$Al in novae/novae-like sources. By comparing the observed 
$^{26}$Al/$^{27}$Al ratio in V4332 Sgr with that expected in classical novae 
it appears unlikely that the progenitor of V4332 Sgr is an 
Oxygen-Neon-Magnesium (ONeMg) white dwarf.

\end{abstract}




\section{Introduction}
      
   $^{26}$Al is an important radionuclide for tracing and studying 
    galactic radioactivity.  $^{26}$Al - the isotope of $^{27}$Al - is
    unstable  with a mean life of $\tau$$_{26}$ = 1.05x10$^{6}$  years
   and decays to $^{26}$Mg emitting a gamma ray photon of 1.809 MeV. It is 
   through the detection of this 1.8 MeV $\gamma$ ray photon
   that  $^{26}$Al  manifests itself.  The  discovery of  $^{26}$Al towards
    the galactic center - deemed to be epochal in nature - was made  by the
    High Energy Astronomical Observatory (HEAO-C) in  1979-80 (Mahoney et al. 
    1982) and  subsequently confirmed by a few other missions (Prantzos $\&$ 
    Diehl 1996). However,    these  missions were limited by inadequate 
    spatial resolution    ( typically a few tens of  degrees) and could 
    not therefore locate the precise sites of the 1.8 MeV $\gamma$ ray 
    emission. A  major advance in mapping the 1.8 MeV $\gamma$ ray emission
    from the galaxy  emerged from recent    studies  by the Compton Gamma ray 
    Observatory (CGRO) (Diehl et al. 1995). The CGRO  mapping  at 
    1.8 MeV - obtained  at a vastly improved resolution (telescope angular
     response is $\sim$    3.8 degrees) but which is still below that 
   required for a precise point source association - traces $^{26}$Al emission 
   in localized regions lying mostly along the galactic plane. A diffuse and
   irregular emission along the galactic plane is observed favoring  
   a massive star origin. Supernovae, novae, Wolf Rayet stars and AGB 
   (Asymptotic Giant Branch) stars are considered the prime candidates for the
  origin of the 1.8 MeV emission but this has  to be established more firmly. 
   Hence a  convincing detection of $^{26}$Al in any of these class of objects
   would give more definite evidence on the exact sites of $^{26}$Al 
   production.  Further, a detection   of $^{26}$Al would  have important 
   implications for current models of nucleosynthesis. Such models give 
   estimates - which are believed to be  uncertain -  of the
   $^{26}$Al/$^{27}$Al ratio    that can   be expected    in the different
   classes of $^{26}$Al producing sources. An observational constraint on 
   the $^{26}$Al/$^{27}$Al  ratio  would therefore  be vitally important for
   nucleosynthesis modeling. The efforts from CGRO are being extended by the
   gamma ray satellite   INTEGRAL which has high-resolution gamma ray 
   spectroscopy as one of its main goals. Thus the study of the 
   origin and  distribution of $^{26}$Al in the galaxy, though in a nascent 
   stage,  is    assuming    significant   importance. It may also be mentioned
   that $^{26}$Al plays an important role in solar system studies where its
   decay to $^{26}$Mg is invoked to explain the excess $^{26}$Mg found in
   meteoritic samples  (the $^{26}$Mg anomaly). It has been proposed that
   the $^{26}$Al and several other short-lived nuclides present in the early
   solar system were freshly synthesized nuclides injected into the proto-solar 
  cloud from a  nearby source such as a nova/supernova/AGB star ( Cameron 
  et al. 1995; Arnould et al.  1997; Goswami $\&$ Vanhala 1998). Alternatively, the possibility of   energetic-particle interactions - as a source of $^{26}$Al production -  has   also been proposed (Gounelle et al.  2001).

The  possibility of a first, spatially-resolved and alternate method of 
detecting $^{26}$Al (apart from detecting the 1.8 MeV line) in an astronomical 
object suggested itself when several strong A-X bands of the AlO radical were 
detected in V4332 Sgr (Banerjee et al. 2003). By virtue of having 
undergone a nova-like outburst, V4332 Sgr satisfied the requisite  criterion 
of belonging to a class of objects  predicted to be a likely site
of $^{26}$Al creation. It was also recognized that molecular bands, arising
from different isotopic compositions, can be easily resolved  spectroscopically
at an intermediate resolution. This is not the case for atomic lines where the
isotope shifts are  too small to detect. A frequently encountered example, of 
isotopically differing molecular bands being resolved, arises in the first 
overtone bands of carbon monoxide at 2.3$\mu$m. Here, the signatures of 
the  $^{12}$CO and $^{13}$CO bands are routinely seen to be distinctly 
separate in the spectra of cool stars. We thus felt that the two components 
of AlO viz. the radioactive component $^{26}$AlO (in case it is present at
detectable limits) and the stable component $^{27}$AlO could be expected to be 
distinctly resolved in V4332 Sgr. Considering that Al bearing molecules are 
seen very rarely in astronomical sources, V4332 Sgr presented a unique chance
to detect the presence of $^{26}$Al. Further, since the AlO
bands in V4332 Sgr are strong, even a low strength of $^{26}$Al vis-a-vis 
$^{27}$Al in the object would still permit the $^{26}$AlO bands to be
detected.  The detected strength of  the putative $^{26}$AlO bands  
would help to establish the $^{26}$Al/$^{27}$Al ratio in the 
source.  While all  the A-X bands of AlO are expected to exhibit
the $^{26}$AlO component,the (2,0) band was chosen for the present 
study because it is strong      and the adjacent spectral region 
around the wings of the band - where  the $^{26}$AlO signature is 
expected - is   relatively  uncontaminated with  other  features. 
Further, the  wavelength separation between the $^{26}$AlO and
$^{27}$AlO components of the  (2,0) band is  larger than that for 
many of the other bands and therefore expected  to be more easily 
resolved.    Specifically, the expected band-origin for the
$A$$^{\rm 2}$$\Pi$$_{\rm 3/2}$  components for $^{27}$AlO and 
$^{26}$AlO are at  1.50351$\mu$m and 1.50155$\mu$m respectively ( a 
separation of 19.6{\AA}); for the  $A$$^{\rm 2}$$\Pi$$_{\rm 1/2}$ 
component the positions are at 1.47492$\mu$m and 1.47325$\mu$m ( a separation 
of 16.7{\AA}) respectively. Such  separations can be  resolved at
a spectral resolution  of 1000 or higher (we have used a resolution
of 3800 in our observations).   

	 It was also felt that the 
	present study could give  additional insights into the outburst 
	mechanism for V4322 Sgr. V4332 Sgr is intrinsically a very
	interesting source having a  rich  optical and near-IR 
	spectrum (Banerjee et al. 2003; Banerjee $\&$ Ashok 2004).
	It  erupted in 1994 with a $\sim$ 9.5 magnitude brightening in the
	$V$ band followed by a slow decline. The distance to the object has been
	estimated to be 300 pc (Martini et al. 1999). The puzzling aspect of 
	the object  was its  rapid  cooling to a M
 giant/supergiant which was 
	uncharacteristic of a classical nova (Martini et al. 1999).
        Classical novae generally evolve towards the hot nebular/coronal state.
        A similar behavior  of evolving towards cool temperatures like 
        V4332 Sgr, has been seen  recently in the spectacular  outburst the 
        eruptive variable V838 Mon  (Munari et al. 2002;  Banerjee $\&$ Ashok,
        2002a; Bond et al. 2003 - for the striking light echo around the
        object). It has been suggested that V838 Mon, V4332 Sgr 
        and  M31 RV (another red variable that erupted in M31; Rich et al. 
        1989) could belong to a new class of objects with outburst properties 
        significantly different from those of conventional eruptive variables 
        like classical, symbiotic, recurrent novae or born again AGB stars 
        (Munari et al. 2002). 
        There is an imperative need to understand the cause of the
 explosion
        in such eruptive variables. Plausible explanations for the outburst
        invoke merger of stars (Soker $\&$ Tylenda 2003 ) and a star
        capturing its encircling planets (Retter $\&$ Marom 2003). The 
        conventional scenario for  outbursts in classical novae - that of a 
        thermonuclear runaway on a white dwarf surface accreting matter from
        its secondary companion - does
        not appear to hold (Bond et al. 2003) but a  variant of it
        may  be applicable to V4332 Sgr given the uniqueness of the
        object. Since in a classical novae outburst, theoretical estimates
        indicate that the  $^{26}$Al/$^{27}$Al ratio  can be  significantly 
        enhanced   to values even exceeding unity , (Gehrz et al. 1999 and 
        references therein; Jose $\&$ Hernanz 1998; Prantzos $\&$ Diehl 1996) 
        the 
        detected strength  of  $^{26}$Al relative to $^{27}$Al in V4332  Sgr 
        could shed  further light on the progenitor of  V4332 Sgr which
        underwent  the  outburst.

        \section{Observations}
           
        $H$ band spectra were obtained on 22 September 2003 with
        the 3.8m United Kingdom IR Telescope (UKIRT), Hawaii and the 
        1-5 micron Imager Spectrometer (UIST) using a 1024$\times$1024 InSb 
        array.  Observations were done by nodding the telescope on two positions
        separated by $\sim$ 12$\arcsec$ along a 2 pixel (0.24$\arcsec$) wide
        slit  resulting in a spectral resolution of 3800.  Flat fielding was
        done with a black body mounted inside UIST and spectral calibration was 
        done using an Argon lamp.  The standard star BS 6998 was observed at 
        similar air-mass as V4332 Sgr and its spectrum was used for ratioing
        the object spectra for removal of  telluric lines. Table 1 gives the 
        details of the spectroscopic observations.

\section{Results}
The  observed spectra in the two dithered positions are 
shown in Figure 1. We show these 
spectra separately to bring out the genuineness and repeatability of several 
finer features that are evident in the bands. The 
spectra here may be compared with the lower-resolution spectrum (R = 450)
of the same bands as given in Banerjee at al. (2003) -  more detail is 
discernible in the present spectra.  The AlO bands described here arise from
transitions between the  A-X electronic states  of the AlO radical (Launila $\&$ 
Jonsson 1994). The A state is a doublet with the spin components 
$A$$^{\rm 2}$$\Pi$$_{\rm 3/2}$   and   $A$$^{\rm 2}$$\Pi$$_{\rm 1/2}$. The
observed spectrum here is typical  of  electronic-vibration transitions in
diatomic molecules (Herzberg 1950). The fine structure seen in the spectra 
arise from the rotational lines  of the P, Q and R branches. The R branch is
manifested most clearly since it gives rise to the prominent band-head 
(blueward of the band origin and marked in Figure 1) where the  R branch 
lines crowd together. The actual structure of the the R band-head is complex
as it consists of two or more branches (Launila $\&$ Jonsson 1994). 
It is around this position of the R branch band-head 
that the contribution of $^{26}$AlO bands is expected to be seen. Since  
hints of an additional component were seen at around this 
position 
in  our earlier data, we  felt that it could be a 
tentative signature for $^{26}$AlO. Specifically we refer to the $J$ 
band spectrum for the $A$$^{\rm 2}$$\Pi$$_{\rm 1/2}$ component of the (4,0)
band  which shows a sloping blue wing at 1.2256$\mu$m
(Banerjee at al. 2003; the blue wings of all the other AlO bands rise 
sharply - typical of molecular band structure). A 
similar $J$ band spectrum for the (4,0) band taken at a higher resolution of 
2050, but not included in Banerjee et al. (2003), decomposes the sloping blue 
wing into two distinct components.   Thus, it was felt that 
the potential  existed to try and make a more detailed search for $^{26}$AlO
in V4332 Sgr.

Because of the good  reproducibility in the two spectra of Figure 1, we have
averaged them  to increase the S/N  and the resultant spectrum is shown 
in Figure 2 (top panel) by the grey line. A computed model spectrum for
AlO - assuming a pure $^{27}$AlO component and no $^{26}$AlO - for the (2,0)
band  for a rotational temperature of 200K is overlaid on this. As may be 
seen, there 
is a good match between the model and the observed data. The broader features, 
like the excellent wavelength matching of the R branch band-head and the
overall band structure are well reproduced. In addition, many of the finer
features due to the individual rotational transitions are well matched.
The observed AlO spectrum of V4332 Sgr appears to be quite pure. However, there
is a broad observed feature at 1.526$\mu$m - marked with an asterisk - which 
does not match the model spectrum and which  appears to be due to some other 
species apart from AlO. The synthetic AlO spectra were created starting from 
the  known molecular constants of both states involved in the transition of 
the experimentally known $^{27}$AlO isotopomer.  Corrections were subsequently 
applied on all these constants, according to standard procedures for isotopic 
corrections, and corresponding $^{26}$AlO spectra were generated using a 
synthesis routine for molecular spectra (available at 
 http://kurslab.physics.kth.se/~olli/superb/).

In the lower panel of  Figure 2, we show four plots which 
indicate  how  the model spectrum for pure $^{27}$AlO would be affected by
including an  $^{26}$AlO component. The  fractional contribution of $^{26}$Al
in each plot is indicated in terms of  the $^{26}$Al/$^{27}$Al ratio 
used to construct each curve. To avoid crowding, we do not show the pure
$^{26}$AlO spectrum - it
has a structure essentially similar to the $^{27}$AlO spectrum but laterally 
shifted to a slightly shorter wavelength (the relative displacements of the
band origin are 19.6 and 16.7{\AA} for the (2,0) doublets respectively). From 
the plots of the lower panel of Figure 2, it is seen that the most distinctive
signature for the presence of $^{26}$AlO would be its R branch band-head labeled 
as feature C. At higher values of the $^{26}$Al/$^{27}$Al ratio (0.3,0.5 and 1)
it is predicted to be strong. Even at a value as low as  0.1, this feature is 
discernible in the computed plots for  the  $A$$^{\rm 2}$$\Pi$$_{\rm 3/2}$ and 
$A$$^{\rm 2}$$\Pi$$_{\rm 1/2}$ components. Hence there should have been some
sign of it in the observed spectra if $^{26}$Al was present at significant 
strength in V4332 Sgr .  The intensity of the feature B with respect to  A, also
 increases with increasing values of the
$^{26}$Al/$^{27}$Al ratio.  The  ratio of these two features in the observed 
spectra would appear to be  best replicated for a  $^{26}$Al/$^{27}$Al ratio 
of 0.1 or lower. Thus, it would appear that $^{26}$Al is not seen at 
appreciable  strength   in V4332 Sgr.  To get a more rigorous 
estimate on the upper limit on the strength of $^{26}$Al, we  applied
a Kolmogorov-Smirnov (KS) test to the observed and model data (for different
$^{26}$Al/$^{27}$Al values). The KS test was applied to a restricted portion 
of the data between 1.47$\mu$m to 1.48$\mu$m. This
section of data was chosen because the strength of features  A, B and C are
expected to be considerably influenced by the presence of $^{26}$Al. The KS
test, in this region,  should therefore be sensitive to the signature
of $^{26}$Al. From the analysis we find a high probability of association
(P) between the model and observed data when $^{26}$Al is not present
(i.e. P = 0.91 for $^{26}$Al/$^{27}$Al = 0). As the  $^{26}$Al/$^{27}$Al 
ratio is 
increased to values of 0.05, 0.1, 0.15 and 0.20 respectively,  P falls to 
values of 0.80, 0.20, 0.033 and 0.001 respectively. The computed probability
values indicate that  $^{26}$Al is definitely not present at large strengths of
15-20 percent. An upper limit of $\sim$ 10 percent is suggested based on the
analysis. The present data do not permit setting a lower limit on the 
strength of $^{26}$Al in V4332 Sgr. Spectra obtained at a higher resolution 
than the present studies - and at similar or greater S/N - where the 
individual rotational lines are better resolved could give a more definite
answer. 

\section{ Discussion }
A positive detection of $^{26}$Al would have constituted definite 
observational confirmation  of the theoretical prediction that novae explosions 
synthesize $^{26}$Al. Even otherwise, the fairly reliable limit of the 
$^{26}$Al/$^{27}$Al ratio that we obtain has important implications. The 
nucleosynthesis of $^{26}$Al in novae and other $^{26}$Al producing sources 
is a very complex field. It is recognized that  considerable uncertainty 
exists in crucial areas related to nuclear reaction networks and nuclear 
reaction rates  thereby leading to uncertainties in the predicted value of the
$^{26}$Al/$^{27}$Al ratio for the different sources (Prantzos $\&$ Diehl 1996;
Gehrz et al. 1998; Busso, Gallino $\&$ Wasserburg 1999). As an example we
 note that the reaction rates for the  key reactions for the production and 
destruction of $^{26}$Al viz. $^{25}$Mg(p,$\gamma$)$^{26}$Al and  
$^{26}$Al(p,$\gamma$)$^{27}$Si (proton capture on $^{25}$Mg and $^{26}$Al) 
have been revised by a factor of 5 in recent times (Prantzos $\&$ Diehl 1996). 
There are additional uncertainties involving  the treatment of convection in 
novae. An observational constraint on 
the $^{26}$Al/$^{27}$Al ratio would therefore be extremely important for 
nucleosynthesis models. The predicted model values for this ratio for 
classical novae are as 
follows. For CO novae (the outburst occurs on a Carbon-Oxygen white dwarf (WD)
with M$_{WD}$ $\leq$ 1.2M$_{\odot}$), the $^{26}$Al/$^{27}$Al ratio depends 
strongly on the WD mass. For WD masses between  0.6 to 1.2M$_{\odot}$, the 
recent results of Starrfield et al. (1997) show that this ratio varies between 
0.01 to 0.6. For the ONeMg novae (which comprise $\sim$ 30-40
percent of all classical novae) the WD is more massive ($>$ 1.2M$_{\odot}$) and
a much greater ratio for $^{26}$Al/$^{27}$Al between 1-10 can be expected 
(Starrfield et al. 1997; Prantzos $\&$ Diehl 1996).  The detailed 
nucleosynthesis models by Jose $\&$ Hernanz (1998) indicate that the
$^{26}$Al/$^{27}$Al ratio can lie between 0.2-0.4 for ONe novae and 
between 0.15-0.56 for CO novae. The increased yield of
$^{26}$Al in ONeMg novae is essentially due to the existing  magnesium rich
environment in the outbursting progenitors in these objects  ($^{24}$Mg and 
$^{25}$Mg  are the seed nuclei for $^{26}$Al generation). As can be seen a 
significant value of the $^{26}$Al/$^{27}$Al ratio  can be expected in novae. 
Novae are estimated to contribute 0.1 to 0.4M$_{\odot}$ (Jose, Hernanz 
$\&$ Coc 1997) of the approximately 2M$_{\odot}$ of $^{26}$Al present in the 
galaxy (a  
value based on the  HEAO-C and CGRO results). While V4332 Sgr is not a 
typical novae, and the predicted $^{26}$Al/$^{27}$Al ratio for classical novae
cannot be strictly extrapolated here, it must still be noted that it underwent 
a single large outburst like a classical nova.

It would be useful to mention the expected $^{26}$Al/$^{27}$Al
ratio in some of the other potential $^{26}$Al producing candidates. 
 In Wolf Rayet stars this ratio  is 2-5x10$^{-2}$ and in type II supernovae 
it is 5-7x10$^{-3}$ . In AGB stars, it is expected to be a few times 10$^{-3}$ 
if $^{26}$Al is produced only in the Hydrogen burning shell. This  ratio is
 uncertain, but can be larger than unity, in case
hot bottom burning is also considered (Prantzos $\&$ Diehl 1996).
For V4332 Sgr specifically,  the lack of $^{26}$Al at appreciable strength, 
indicates that its  progenitor is unlikely to be a  ONeMg white dwarf. As
mentioned earlier, there is keen interest in understanding the  nature of
the progenitor and the outburst mechanism for this object and its probable analog
V838 Mon.  The present results give additional insights in this direction. It may 
also be noted that  $^{22}$Na (mean life $\sim$ 3.75 years) is another important
radioisotope whose production is expected to be significantly enhanced in
nova outbursts (Starrfield et al. 1997, Gehrz et al. 1998). Emission lines from 
the resonance doublet of NaI at 5890, 5896{\AA} have been detected with
considerable strength in V4332 Sgr (Banerjee $\&$ Ashok 2004). If a  significant 
fraction of the observed NaI emission is from $^{22}$Na, then because of its 
short mean life, a considerable reduction in the intensity of the NaI doublet can 
be expected (vis-vis other emission lines seen in the spectrum) over a timescale 
of a few years. The object should be monitored to detect such changes. $^{22}$Na 
is also of interest for $\gamma$ ray line study by INTEGRAL because of the 
1.275 MeV photon emitted by its decay to $^{22}$Ne. 
      
 	\acknowledgements
 	We thank the anonymous referee for constructive comments.
        The research work at PRL is funded by the Department of Space,
        Government of India. We thank the UKIRT service program for
        observation time  and M.Seigar and O.Kuhn of 
        UKIRT for doing the observations.  UKIRT is operated by
        JAC, Hawaii, USA, on behalf of the UK PPARC.


\clearpage 
\begin{figure}
\plotone{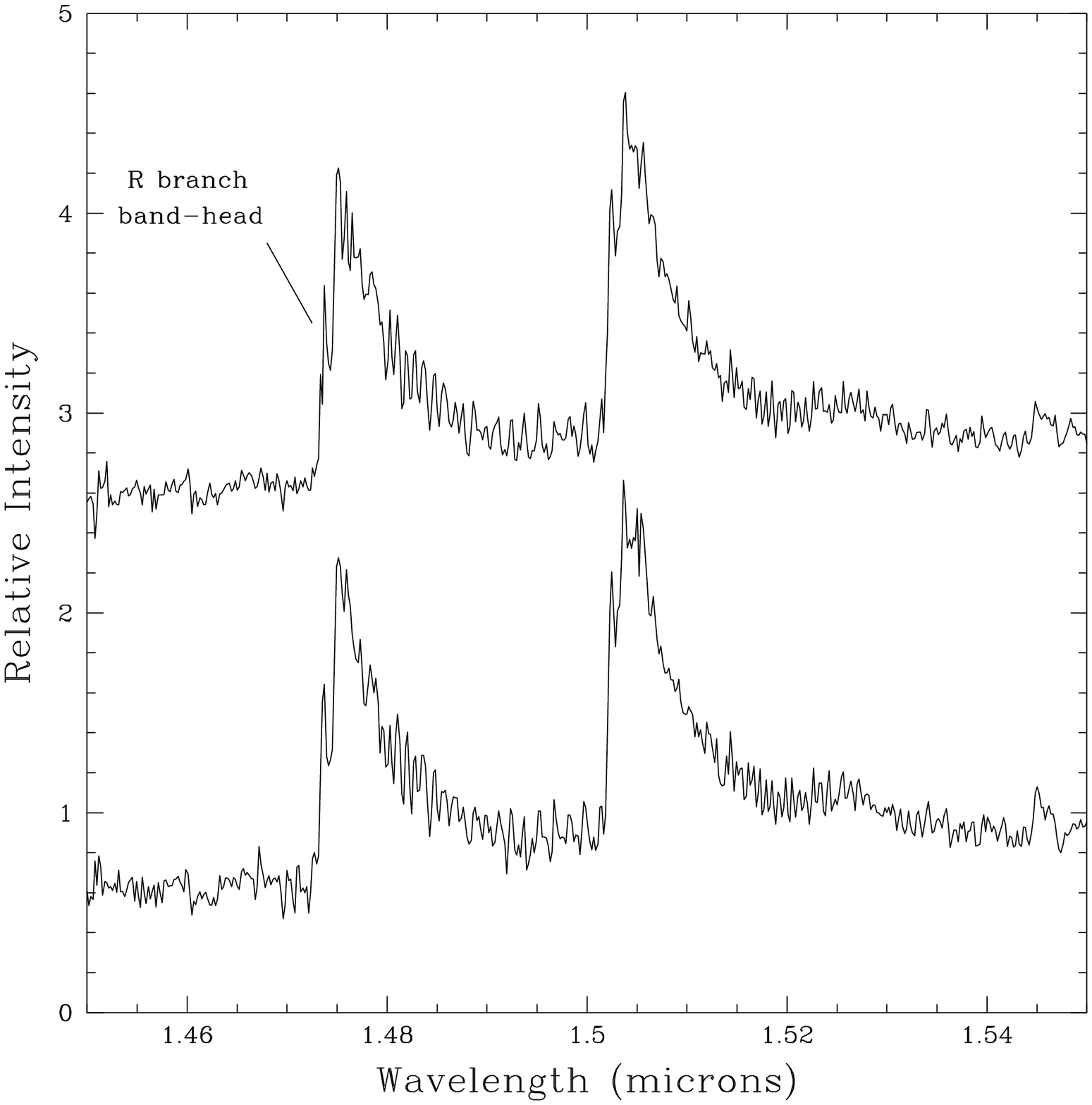}
\caption{The two observed  spectra of V4332 Sgr covering the (2,0) band
of the A-X band system of AlO. The spectra were obtained from UKIRT
on 22 September 2003 at a resolution of $\sim$ 3800.  \label{fig1}}
\end{figure}

\clearpage
\begin{figure}
\plotone{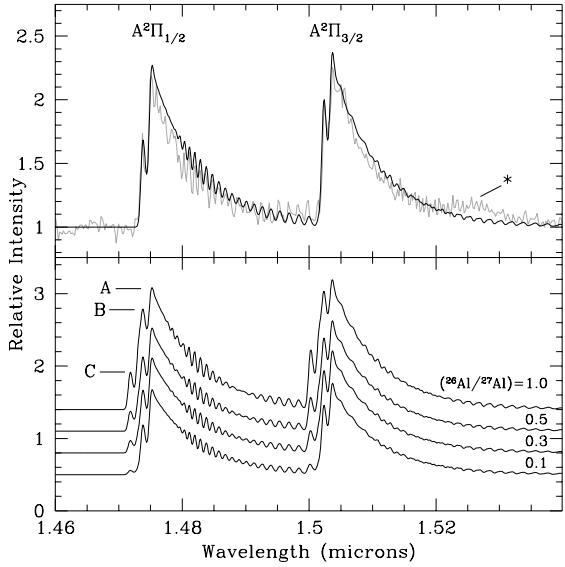}
\caption{The top panel shows  the sum of the two spectra of Figure 1 (in gray)
with a slope correction applied. Overlaid on this is a model spectrum of 
$^{27}$AlO (black line) for an assumed rotational temperature of 200K.  A good
match is found except for the broad feature marked with 
an asterisk which is due to an unidentified   species and  
not AlO. The bottom panel shows the effect on the pure $^{27}$AlO spectrum
of the top panel by including a contribution from $^{26}$Al at different 
strengths. Please refer Section 3 for further details.\label{fig2}}
\end{figure}





\clearpage
\begin{table}
\caption{Log of spectroscopic observations }
\begin{tabular}{ccccc}
\hline \\ 
Object & UT    &  Exposure & Integration & Mean\\
       &       & Time(s)   & Time(s)     & Airmass\\
\hline 
\hline 
V4332 Sgr & 5.558   & 240  & 2880  & 1.35 \\
BS 6998    & 5.067   & 3    & 48    & 1.32  \\
\hline
\end{tabular} 
\end{table}

\end{document}